\documentclass[traditabstract]{aa} 
\usepackage{graphicx}
\usepackage{txfonts}
\usepackage{natbib}
\usepackage{lscape}

\begin{document}

\title{The AKARI NEP-Deep survey: a mid-infrared source catalogue\footnote{Full Tables 4 and 5 are only available 
at the CDS via anonymous ftp to {\tt cdsarc.u-strasbg.fr (130.79.128.5)} or via {\tt http://cdsarc.u-strasbg.fr/viz-bin/...   }}}

   \author{T. Takagi \inst{1}
                    \and
	H. Matsuhara \inst{1} \and
	T. Goto \inst{2,3}  \and
	H. Hanami \inst{4} \and 
		M. Im \inst{5} \and 
	K. Imai \inst{6} \and 
	T. Ishigaki  \inst{4} \and 
	H.M. Lee \inst{5} \and 
	M.G. Lee \inst{5} \and 
	M. Malkan \inst{7} 
          Y. Ohyama \inst{8}  \and
	S. Oyabu \inst{9} \and
          C.P. Pearson \inst{10} \and
         	S. Serjeant \inst{11} \and 		
	T. Wada \inst{1} \and 
	G.J. White \inst{10,11} 
          }

   \institute{
   Institute of Space and Astronautical Science, Japan Aerospace Exploration Agency, 
   Sagamihara, Kanagawa 229-8510, Japan \\
                   \email{takagi@ir.isas.jaxa.jp}
             \and
             Institute for Astronomy, University of Hawaii, 2680 Woodlawn Drive, Honolulu, HI, 96822, USA
             \and
             Subaru Telescope 650 North A'ohoku Place Hilo, HI 96720, USA
                          \and
             Physics Section, Faculty of Humanities and Social Sciences, Iwate University, Morioka 020-8550, Japan
                          \and 
             Department of Physics and Astronomy, FPRD, Seoul National University, Shilim-Dong, Kwanak-Gu, 
             Seoul 151-742, Korea
                          \and
             TOME R\&D Inc. Kawasaki, Kanagawa 213-0012, Japan
                          \and
             Department of Physics and Astronomy, UCLA, Los Angeles, CA, USA
               \and
             Academia Sinica, Institute of Astronomy and Astrophysics, Taiwan 
             \and
             Graduate School of Science, Nagoya University, Furo-cho, Chikusa-ku, Nagoya, Aichi 464-8602, Japan
             \and
             Rutherford Appleton Laboratory, Chilton, Didcot, Oxfordshire, OX11 0QX, UK 
             \and 
             Astrophysics Group, Department of Physics, The Open University, Milton Keynes, MK7 6AA, UK
             }

   \date{Received 22 July 2011; accepted 8 November 2011}

 
  \abstract 
{We present a new catalogue of mid-IR sources using the {\it AKARI} NEP-Deep survey. The InfraRed Camera (IRC) onboard {\it AKARI} has a comprehensive mid-IR wavelength coverage with 9 photometric bands at 2 -- 24\,$\mu$m. We utilized all of these bands to cover a nearly circular area adjacent to the North Ecliptic Pole (NEP). We designed the catalogue to include most of sources detected in 7, 9, 11, 15 and 18\,$\mu$m bands, and found 7284 sources in a 0.67\,deg$^2$ area. From our simulations, we estimate that the catalogue is $\sim80$ per cent complete to 200\,$\mu$Jy at 15 -- 18\,$\mu$m, and $\sim$10 per cent of sources are missed, owing to source blending. Star-galaxy separation is conducted using only {\it AKARI} photometry, as a result of which 10 per cent of catalogued sources are found to be stars. The number counts at 11, 15, 18, and 24\,$\mu$m are presented for both stars and galaxies. A drastic increase in the source density is found in between 11 and 15\,$\mu$m at the flux level of $\sim300$\,$\mu$Jy. This is likely due to the redshifted PAH emission at 8\,$\mu$m, given our rough estimate of redshifts from an {\it AKARI} colour-colour plot. Along with the mid-IR source catalogue, we present optical-NIR photometry for sources falling inside a Subaru/Sprime-cam image covering part of the {\it AKARI} NEP-Deep field, which is deep enough to detect most of {\it AKARI} mid-IR sources, and useful to study optical characteristics of a complete mid-IR source sample. 
}


   \maketitle
%

\section{Introduction}

Over the last two decades, the space infrared missions have made deep extragalactic surveys, which produced a growing sample of infrared-luminous galaxies. When {\it IRAS} discovered a population of ultra-luminous infrared galaxies \cite[ULIRGs -- see][for a review]{1996ARA&A..34..749S}, they were regarded as very rare objects like QSOs. {\it IRAS} also provided good evidence for number density evolution of infrared-luminous galaxies, although the redshift range was limited to $z\la 0.2$ \citep{1987ApJS...63..311H,1990ApJ...358...60L,1990MNRAS.242..318S,1995AJ....110..259G,1997A&A...323..685B}. The next infrared mission, {\it ISO}, provided further evidence of evolution towards $z\sim1$ \citep{1999ApJ...517..148F,1997MNRAS.289..471O,1997MNRAS.289..490R,2000ApJ...541..134X,2004MNRAS.355..813S,2005AJ....130.2019O}. These measurements are mostly summarized in terms of the cosmic star formation rate density, estimated from the integrated infrared luminosity function. Probing deeper and wider areas, more advanced space telescopes, such as {\it Spitzer}, {\it AKARI} and {\it Herschel},  firmly established strong evolution of infrared galaxies, now in the form of evolving infrared luminosity function. Results from various extragalactic survey fields converge specifically in the redshift range $z \la 1$. That is, the comoving infrared energy density evolves as rapid as $(1+z)^4$ \citep{2002A&A...384..848E,2004MNRAS.355..813S,2005ApJ...632..169L,2005ApJ...630...82P,2006MNRAS.370.1159B,2007ApJ...660...97C,2009A&A...496...57M,2010A&A...514A...6G,2010A&A...515A...8R,2010A&A...518L..27G,2011arXiv1101.2467M}. At $z\sim 1$, luminous infrared galaxies (LIRGs) with $L_{IR} > 10^{11} L_\odot$ are responsible for more than half of this infrared energy density \citep[e.g.][]{2005ApJ...632..169L}. This also indicates that infrared-luminous galaxies produced a significant fraction of stellar mass in the present universe. Thus, they were indeed major galaxy population in the past, giving important clues as to how galaxies evolve. Statistical samples of (U)LIRGs are vital for studies of galaxy formation and evolution. 

{\it AKARI} is the first Japanese space mission dedicated to infrared astronomy \citep{2007PASJ...59S.369M}. {\it AKARI} was launched in Feb 2006 with the M-V-8 rocket from the Uchinoura Space Centre in Japan, as a second generation all-sky surveyor at infrared wavelengths. The first point-source catalogues from the all-sky survey were released in March 2010. Along with this all-sky survey, {\it AKARI} conducted 5088 pointed observations in selected areas of sky, during its liquid helium cold phase. 
Using 13 percent of the pointed observation opportunities in the cold phase, we conducted an extragalactic survey around the North Ecliptic Pole (NEP). This NEP survey is two-tiered, consisting of the NEP-Deep and the NEP-Wide survey. A salient characteristic of this survey is its comprehensive mid-IR wavelength coverage -- we used 9 photometric bands to span the wavelength range from 2 to 24\,$\mu$m. The scientific advantage of obtaining such comprehensive wavelength coverage include a capability to reliably distinguish starburst-dominated galaxies from those with AGN contributions, to obtaining rest-frame mid-IR fluxes, free from the uncertainty of complicated $K$-corrections. The NEP survey provides a valuable input to the study of infrared-luminous galaxies. 

This paper describes the mid-IR source catalogue produced with the NEP-Deep survey. The description of the data in the NEP-Wide is given elsewhere \citep{2009PASJ...61..375L,2010ApJS..190..166J}.  We give a brief summary of observations and data reduction in section 2, including those for ancillary observations from the ground. Source identification and photometry are presented in section 3, including performance checks using simulated sources. We describe the generation procedures for the catalogue, including optical identification and star-galaxy separation in section 4, and give some discussion in section 5. Our summary is given in section 6. 

Use of a concordance cosmology\footnote{$\Omega_m = 0.3, \Omega_\Lambda=0.7$ and $H_0 = 70$\, km\,sec$^{-1}$\,Mpc$^{-1}$} gives a scale of 1.8\,kpc/arcsec at $z=0.1$. Therefore, the majority of the galaxies appear point-like at $z\ga 0.1$ at the resolution of {\it AKARI}, that is $\sim 5''$. 

\section{Observations and data reduction}

\subsection{AKARI IRC}

The NEP-Deep survey covers a circular field with an area of 0.67\,square degrees\footnote{Area covered with $L18W$ before any masking.} using the IRC onboard {\it AKARI}, with a field-of-view of $10' \times 10'$. Since details of observations and data reduction are described in \cite{2008PASJ...60S.517W}, we give only a brief summary here. 

The design of {\it AKARI}'s Sun shield and attitude determination system require that the optical axis of the telescope is always kept pointing 90$^\circ$ from the direction to the Sun with a tolerance of only $\pm 0.6^\circ$. Due to this strong visibility constraint in its sun-synchronous orbit, deep surveys are possible only close to the Ecliptic Poles. Given the presence of the Large Magellanic Cloud near the South Ecliptic Pole, we chose the North Ecliptic Pole (NEP) for  {\it AKARI's} unique deep survey field \citep{2006PASJ...58..673M}. 

\begin{figure}
\begin{center}
  \resizebox{6cm}{!}{\includegraphics{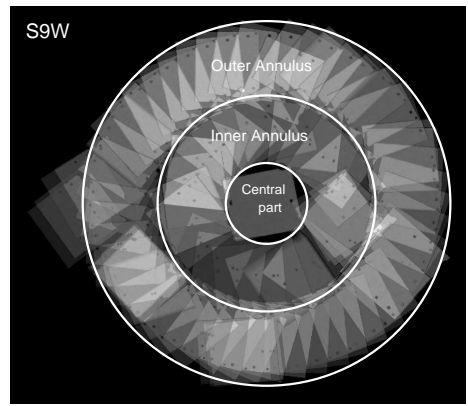}}
\end{center}
 \caption{Coverage map of the NEP-Deep survey at 9\,$\mu$m, showing three regions. Square patterns indicate the field-of-view of 
 the MIR-S channel of the IRC. See text for details. }
 \label{field}
\end{figure}

The circular field of the NEP-Deep survey can be divided into three regions -- the central field,  and the inner and outer annuli as shown in Figure \ref{field}. The size of the inner annulus is determined by the angular separation between the NIR/MIR-S channel and the MIR-L channel\footnote{The NIR and MIR-S channels share the same field-of-view.}. During a pointed observation of the inner annulus, the NIR/MIR-S channels always observe the opposite side of the MIR-L channel. The angular separation between the centers of the IRC channels is 20$'$. This separation and the field-of-view of IRC result in a central hole $\sim$10$'$ in diameter. This central hole defines the central region (RA = 17h56m, Dec = 66$^\circ$37$'$) and was covered with one field-of-view of the IRC. Because of the difference in the field shape, i.e. a square field-of-view for a circular area, some small gaps exist between the central region and the inner annulus. Since {\it AKARI} has no capability to control its roll angle, we utilized the seasonal variation of the position angle to cover the survey field. The radius of the outer annulus was set to 20$'$ so that there would be no gap between the inner and outer annuli. In total we allocated 23, 63, and 146 pointed observations for the central region, the inner and the outer annuli, respectively. We adopted an observing mode with neither filter change nor dithering (i.e. so-called AOT05), which is optimum for deep surveys. Since each channel of the IRC has 3 filters, we need at least 3 pointed observations to cover the entire wavelength range of the IRC. On average, 4 pointed observations per filter were conducted at a given position. 

The observations were executed from 2006 May to 2007 August. Unfortunately, the quality of the resulting mid-IR image depends on the season when the observation was conducted. From April to August, the observations towards the NEP were affected by stray light from the Earth shine. This degradation is severer for longer-wavelength images, and affects the north-east part of the field in an area of $\sim$0.15 deg$^2$. 

We used the standard IRC imaging pipeline version 20071017\footnote{See AKARI IRC Data User Manual ver 1.3} for each pointed observation, in which dark subtraction, subtraction of scattered light inside the camera, correction for detector non-linearity, flat fielding, and correction for distortion are performed. For the MIR-S and MIR-L images, we also removed the diffuse background by subtracting a median-filtered image. Astrometry in all of the near-IR and $S7$ and $S9W$ images was determined by the IRC pipeline using 2MASS objects as a reference. For images at longer wavelengths, we used the coordinates of bright sources in the image one-band shorter in wavelength than the image in question. The individual images of a given band with calibrated astrometry were combined to produce a final mosaicked image by using a publicly available software, Swarp\footnote{Available at {\tt http://www.astromatic.net/}}. The images were combined by taking a median value of the corresponding individual images. Overviews and close-ups of these final images have been presented in Wada et al. (2008). 

Wada et al. (2008) have measured the sky noise limit from photometry of random positions containing no sources. The 5\,$\sigma$ detection limits of a point source are estimated to be 9.6\,$\mu$Jy in $N2$ band, 7.5\,$\mu$Jy ($N3$), 5.4\,$\mu$Jy ($N4$), 49\,$\mu$Jy ($S7$), 58\,$\mu$Jy ($S9W$), 71\,$\mu$Jy ($S11$), 117\,$\mu$Jy ($L15$), 121\,$\mu$Jy ($L18W$) and 276\,$\mu$Jy ($L24$)\footnote{Corrected for the error in the conversion factor used in Wada et al. (2008)}. Hereafter we use these values as the nominal detection limits of the NEP-Deep survey.

\subsection{Ancillary data}
For the NEP field, we performed various follow-up observations from X-ray to radio wavelengths \citep[e.g.][]{2010A&A...517A..54W,2010yCat..35179054W}. Optical-NIR photometry of mid-IR sources in this paper utilizes the following data -- Subaru/Suprime-cam ($BVRi'z'$), CFHT/Megacam ($u^*$), KPNO 2.1m/FLAMINGOS ($JK_s$), which cover $\sim 40$ per cent of the NEP-Deep field, except for the $u^*$ band. An optical source catalogue covering the whole NEP-Deep field with CFHT/Megacam ($g'r'r'z'$) is published in \cite{2007ApJS..172..583H}, to which readers interested in optical properties of more sources are referred. 

\subsubsection{Subaru/Suprime-cam}
We used the Subaru Suprime-cam (S-cam) to obtain deep optical images for a single field-of-view, i.e. $34' \times 27'$, covering 38 per cent of the NEP-Deep field. Observations in the $B, R, i', z'$ bands were conducted in June 2003. Additional $V$-band observations were performed in Oct 2003 and June 2004. The central position of the field (RA=17h55m24s, Dec=$+$66$^\circ$37$'$31$''$ [J2000]) and the position angle (PA=0$^\circ$) were selected to avoid the presence of bright stars ($V<10$\,mag) in the field. 

We obtained integration times of 12960\,sec for the $B$ band, 7421.2\,sec for $V$ band, 7200\,sec for $R$ band, 6900\,sec for $i'$ band, and 10080\,sec for $z'$ band, with a typical seeing of 0.6$''$ -- $1''$. Frames which had seeing worse than 1$''$ were not used in processing the final mosaiced images. The data were reduced in a standard manner using SDFRED \citep{2002AJ....123...66Y,2004ApJ...611..660O}. Sources are extracted with SExtractor \citep{1996A&AS..117..393B} using the $z'$-band image as a detection image. The 5-$\sigma$ limiting magnitudes measured with a 2$''$ aperture were 28.4\,mag for $B$ band, 28.0\,mag for $V$, 27.4\,mag for $R$, 27.0\,mag for $i'$ and 26.2\,mag for $z'$ in the AB magnitude system. 

\subsubsection{CFHT/Megacam ($u^*$)}
We obtained an UV image of the whole NEP-Deep field with CFHT/Megacam. Observations in the $u^*$ band were taken in queue mode spread over 12 nights from April 2007 to September 2007, resulting in a total of 77 frames. The field of view of Megacam was large enough to cover the entire NEP-Deep field with the central position of RA 17h55h24s and Dec  $+$66$^\circ$37$'$32$''$ [J2000]. With the integration time of 600\, sec for each frame (2 frames have 680\,sec), we achieved the total integration time of 773\,min. Most of these data--56 frames out of 77--were taken at airmasses less than 1.5 and typical seeing of 1.0$''$. 

The data reduction was carried out in a standard manner, using software developed for the large-format Megacam data. The pre-processing, including bad-pixel masking, overscan and bias subtraction, and flat-fielding, were carried out before the data delivery using the pipeline system `Elixir'. By stacking the Elixir-processed images, we produced a final mosaiced image using a software package provided by TERAPIX\footnote{http://terapix.iap.fr} including WeightWatcher, SExtractor, SCAMP and SWarp. The depth of the final image was estimated to be 24.6 mag [5\,$\sigma$; AB] from photometry of 30000 random positions with 2$''$ aperture.


\subsubsection{KPNO 2.1m/FLAMINGOS}
\cite{2007AJ....133.2418I} describe observations with KPNO\,2.1m FLAMINGOS and its data reduction in detail, and therefore we will give only a brief summary here. We conducted FLAMINGOS observations in the $J$ and $K_s$ bands to cover the S-cam field in the NEP-Deep field. We covered the target area with observations at 4 different positions. The total effective area is 750\,arcmin$^2$, with stellar image sizes ranging from 1.1$''$ to 2$''$ in FWHM. Data were reduced in the standard manner using the IRAF packages. The deepest regions reach $J=22.2$ and $K_s=21.3$ AB magnitude in 5\,$\sigma$. 

 \begin{figure*}
\begin{center}
  \resizebox{8cm}{!}{\includegraphics{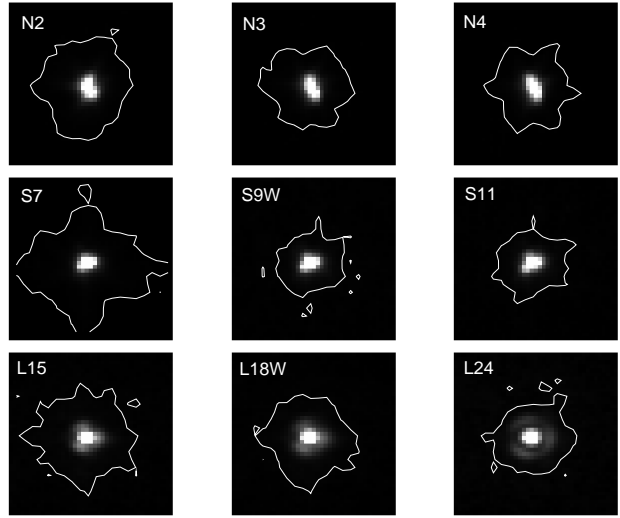}}
\end{center}
 \caption{Average PSFs obtained with stacking of bright sources, displayed with a linear grey scale covering 99\,\% of the dynamic range. Solid curves indicate the area with signal-to-noise ratio of 5 or greater. }
 \label{psf}
\end{figure*}

 \begin{figure*}
\begin{center}
  \resizebox{10cm}{!}{\includegraphics{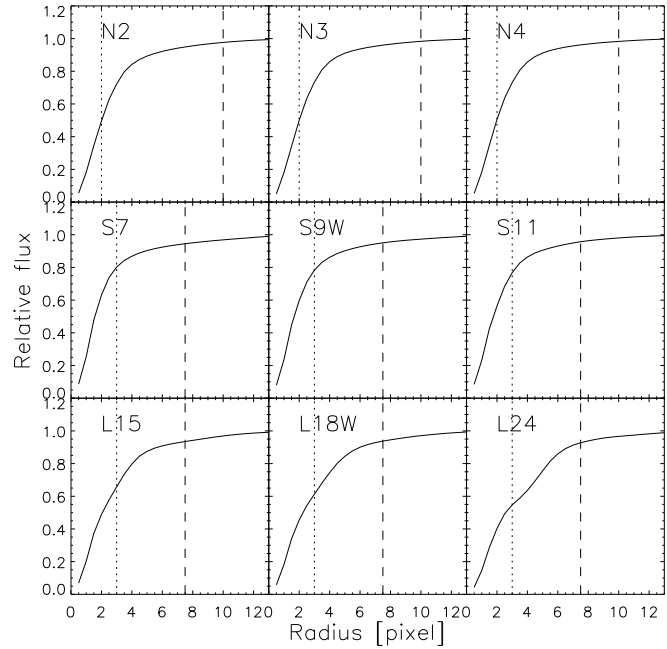}}
\end{center}
 \caption{Growth curves of enclosed fluxes as a function of aperture radius. The vertical axis reports relative enclosed fluxes, normalized 
 to the value at a 15-pixel radius. Dotted and dashed lines indicate adopted apertures for photometry and those 
 for flux calibration in \cite{2008PASJ...60S.375T}. 
} \label{gcurve}
\end{figure*}

\section{Source identification and photometry}

\subsection{Source detection in MIR-S and MIR-L}

We combined the $S7$, $S9W$, and $S11$ images with the terapix software 'SWarp' to produce a detection image for the MIR-S channel, median-coadding three separate images. For the MIR-L, we used the average of the $L15$ and $L18W$ images for source detection, since the quality of $L24$ image was systematically lower. In the resulting MIR-S and MIR-L detection images, the residual from the sky subtraction was reduced significantly, which provided more reliable source catalogs. 

For each detection image, we ran SExtractor \citep{1996A&AS..117..393B} to produce initial source lists of mid-IR sources. In order to make these initial source lists as complete as possible, we adopted a relatively low threshold for source detection, i.e. requiring 5 connected pixels with $>$1.2\,$\sigma$. This resulted in 6746 MIR-S sources and 6719 MIR-L sources. Mainly because of the background fluctuation and the low threshold, these may include a number of spurious sources. Thus, these sources were regarded as `candidates', and are screened with criteria described below. In the final catalogue, the MIR-S and MIR-L catalogues were concatenated where duplicated entries are removed.

\begin{table*}
\begin{center}
\caption{Summary of point spread functions of IRC}
\begin{tabular}{l|ccc|ccc|ccc}
\hline
\hline
                                 &  N2  &   N3   &   N4 &   S7 &   S9W &   S11 &   L15 &   L18W &   L24 \\
\hline
 Reference wavelength [$\mu$m]  & 2.4  & 3.2 & 4.1 & 7.0 & 9.0 & 11.0 & 15.0 & 18.0 & 24.0 \\
\# of postage stamps used $^a$ &168&114&43&45&28&24&56&66&43 \\
FWHM (pixel)$^b$ & 2.96 & 3.06 & 2.84 & 2.27 & 2.19 & 2.10 & 2.27 & 2.48 & 2.92 \\
Aperture radius (pixel) & 2 &2 &2 & 3& 3  &3&3&3&3 \\
Aperture correction factors$^c$ &  1.97 & 1.96 & 1.95 & 1.18 & 1.21 & 1.25 & 1.42 & 1.53 & 1.70 \\
Pixel scale & \multicolumn{3}{|c|}{$1.46\times 1.46$} & \multicolumn{3}{c|}{$2.34\times 2.34$} &\multicolumn{3}{c}{$2.51 \times 2.39$} \\
\hline 
\multicolumn{10}{l}{$^a$ The number of stacked postage stamp images for PSF creation} \\
\multicolumn{10}{l}{$^b$ FWHM computed using the enclosed flux radial profile } \\
\multicolumn{10}{l}{$^c$ Correction factors for fluxes }\\
\end{tabular}
\end{center}
\label{tb::psf}
\end{table*}

\subsection{Photometry and band-merging}
We followed the procedures of Takagi et al. (2007) for photometry and band-merging for each MIR channel. Starting from the initial source position obtained with SExtractor, we searched the centroid position for sources in each IRC image. The re-evaluated centroid position could shift from the SExtractor position significantly in case of severe source blending or low signal-to-noise ratio. Therefore, we performed photometry only if the new centroid is less than 3 arcsec away from the initial position,  which corresponds to $\sim$2\,$\sigma$ of the relative offset between MIR-S and MIR-L sources.

We performed aperture photometry in each IRC image at the centroid positions determined above, adopting a 2-pixel ($2.9''$) radius for NIR images and 3- pixel ($7.0''$) radius for MIR-S and MIR-L images. We then applied the aperture corrections, estimated from the average digitized point spread function (PSF) created from stacking of bright sources around the NEP. In this stacking, we did not apply any weight as a function of fluxes, since high weight in bright sources also results in high weight for associated sources in the outskirts. This reduced the signal-to-noise ratio of the outer part of the PSF, and increased the uncertainty of the aperture correction. We show the average PSFs and the growth curves in Figure \ref{psf} and \ref{gcurve}, respectively. From these growth curves, we derived the aperture correction factors, which are tabulated in Table 1 along with other characteristics of the PSFs. The aperture-correction factors derived here are defined as the ratio of fluxes with two different aperture sizes, i.e. the small apertures adopted here and the IRC calibration apertures\footnote{In calibrating the IRC photometry, \cite{2008PASJ...60S.375T} compared the aperture photometry with the expected total fluxes from spectral models of calibration stars. Therefore, the IRC calibration factors are supposed to include the aperture correction from the calibration aperture photometry to the total fluxes. } \citep[10 pixels for the NIR channel and 7.5 pixels for the MIR channels, see][]{2008PASJ...60S.375T}. The background level was estimated in the sky annulus between a 15- and 20-pixel radius. 

Although this photometric method is simple, it is relatively robust against the effects of source blending, compared to photometry with larger apertures adopted in the IRC calibration or MAG\_AUTO photometry in SExtractor. A similar approach was successfully applied to photometry of blended sources in previous works \citep[e.g.][]{2009ApJ...695L.198K}. In heavily confused images, photometry with simultaneous PSF-fitting for multiple sources may be a better solution. However, there are arguments against PSF fitting in our case. It is difficult to accurately determine the PSF in each frame, due to uncertainty in the pointing stability and the lens aberration whose direction is fixed in the detector array. 
This hampers rigorous PSF-fitting analysis. Moreover, faint sources have only a few significant pixels that are useful for fitting. Therefore, we adopted aperture photometry, rather than PSF-fitting photometry. 

\begin{figure*}
\begin{center}
  \resizebox{12cm}{!}{\includegraphics{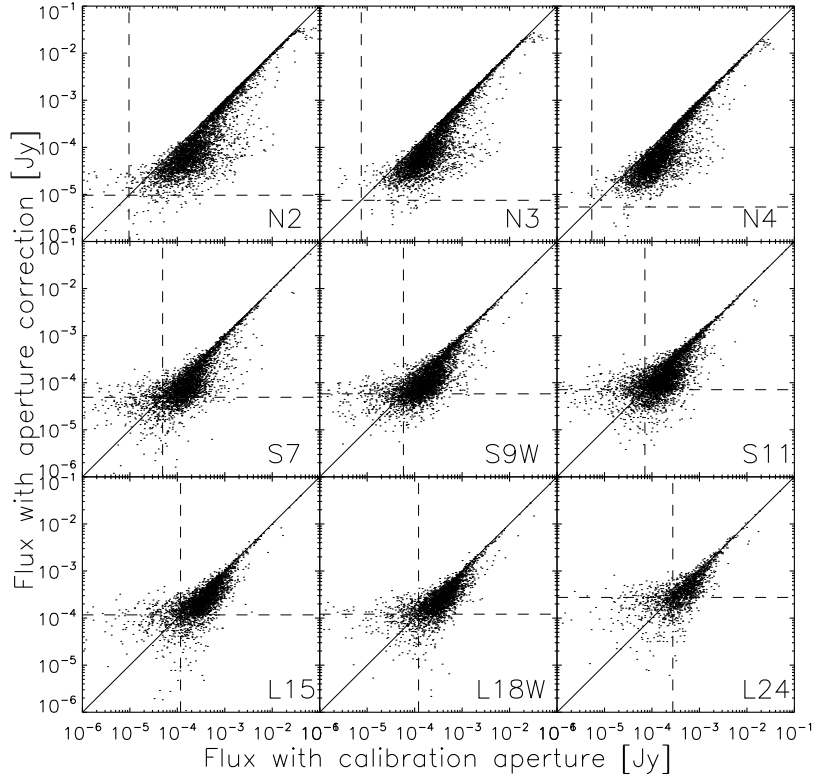}}
\end{center}
 \caption{Fluxes with aperture correction adopted here versus those using the IRC calibration apertures, i.e. 10 pixels for NIR and 7.5 pixels for MIR-S and MIR-L. Dashed lines indicate 5\,$\sigma$ sensitivity in Wada et al. (2008). Solid lines correspond to the linear relation. 
}
 \label{phcomp}
\end{figure*}

Figure \ref{phcomp} compares aperture-corrected fluxes to those obtained with the aperture photometry using the aperture size adopted for the IRC calibration, i.e. 10 pixels for NIR and 7.5 pixels for MIR-S and MIR-L. For bright sources, these fluxes are consistent with each other, except for the brightest ones that have saturated pixels. This confirms that the aperture correction obtained from the average PSFs is reasonable. On the other hand, fluxes of fainter objects, but detected with $\sim 10\,\sigma$, show systematic deviation -- fluxes with the calibration aperture are systematically larger. For fainter sources, the number density increases, and therefore the probability that neighbouring sources fall in the large aperture for IRC calibration increases as well. These associated sources would boost the fluxes with large aperture size as evidently seen in Figure \ref{phcomp}. 

\subsection{Astrometry}
\begin{figure*}
\begin{center}
  \resizebox{12cm}{!}{\includegraphics{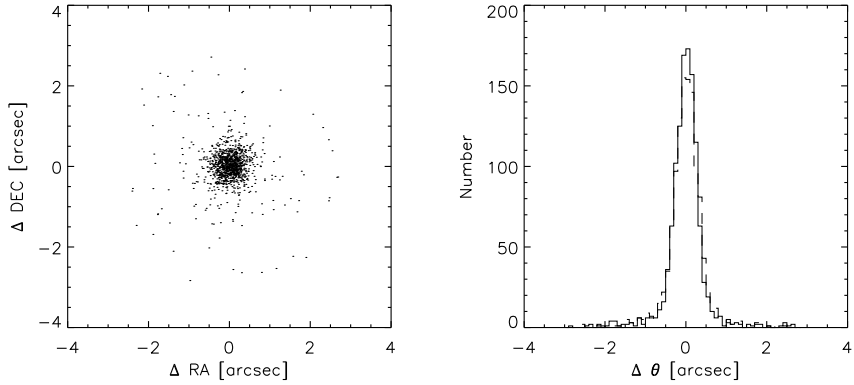}}
\end{center}
 \caption{Comparison of IRC coordinates with the 2MASS coordinates. In the histogram plot, solid and dashed lines indicate $\Delta$RA and $\Delta$Dec, respectively. 
}
 \label{astcheck}
\end{figure*}

The astrometric calibration of IRC images has already been discussed by  Wada et al. (2008), and will not be repeated here.
We adopt the coordinate at the shortest wavelength band detected as the final IRC coordinate of MIR sources. In Figure \ref{astcheck}, we compare the IRC coordinates with 2MASS coordinates for 1178 sources. The mean and $\sigma$ of the coordinate shift are $-0.0014''$ and 0.29$''$ for Right Ascension, and 0.011$''$ and 0.32$''$ for Declination, respectively. 

\subsection{Simulations with artificial sources}
In order to estimate the photometric uncertainties and completeness of our catalogue, we made a careful simulation to estimate this.
Artificial sources were randomly added to the final NEP-Deep images and photometry was done in the same 
manner as for real sources. We consider only point sources here, since the source detection and photometry are optimized for such sources. 
We used the PSFs shown in Figure \ref{psf} for every artificial source, adding 200 artificial sources (less than $\sim$3 per cent of true sources) at a time, and repeating this for a total of 40,000 and 160,000 artificial 
sources, to estimate the photometric errors and completeness, respectively.

\subsubsection{Photometric errors}
We performed aperture photometry for artificial sources as was done for the real sources, and then applied the aperture 
corrections. The centroid positions of the artificial sources were determined through Gaussian-fitting of the simulated point source image, using the input position as an initial guess. Thus, the effect of noise on the source position was included in the simulation. In Table 2, we have tabulated the relative flux errors estimated from the input-to-output flux ratio. We note that the artificial sources were added to the images with real sources, and therefore the photometric errors reported here have some contribution from the source blending. Here the photometric errors were calculated from the scatter of the output fluxes with the rejection of outliers, which are due to the presence of nearby real bright sources.  Although the catalogue generation method is optimized for point sources, the catalogue includes nearby galaxies, for which IRC fluxes are not as accurate as for point sources. We also caution that near-IR fluxes of the several brightest stars are saturated and therefore not accurate. 

\subsubsection{Completeness and reliability}
The completeness of the resulting catalogue is estimated again with the artificial-source method. To reduce the Poisson error of the completeness, we produced a large number of artificial sources--over 160,000. Since we combine multi-band images for source detection, the completeness depends on the colour of sources. We consider two cases for the MIR-S; a) flat spectrum in $f_\nu$ and b) red sources, for which we adopt the following flux ratios: $f_{9\mu m}/f_{7\mu m} = 6$ and $f_{11\mu m}/f_{7\mu m} = 10$ in Jansky units.  This corresponds to the reddest colours among the sources detected in all of MIR-S bands.\footnote{Our main concern for the source extraction procedure was the effects of combining three MIR-S images taken at  the different wavelengths. In this procedure, sources with extreme colours may not be detected in some of MIR-S bands. In order to evaluate conservatively such effects, we conducted the simulation with the most extreme colour.} The average flux ratios of MIR-S--detected sources were found to be $\langle f_{9\mu m}/f_{7\mu m} \rangle = 1.8$ and $\langle f_{11\mu m}/f_{7\mu m}\rangle = 2.6$. For the MIR-L channel, only the flat spectrum case was considered, because we only combined 15 and 18\,$\mu$m images where most of galaxies have similar fluxes. 

We ran SExtractor on each simulated image with 200 artificial sources. A source was considered to be recovered if its extracted position was within a radius of 3$''$. In some other studies, the consistency of recovered fluxes is also taken into account as a condition of the source recovery, in order to avoid mis-identification with real sources. In our case, this flux check is not straightforward, since the detection images we made are not calibrated.

\begin{figure}
\begin{center}
  \resizebox{8cm}{!}{\includegraphics{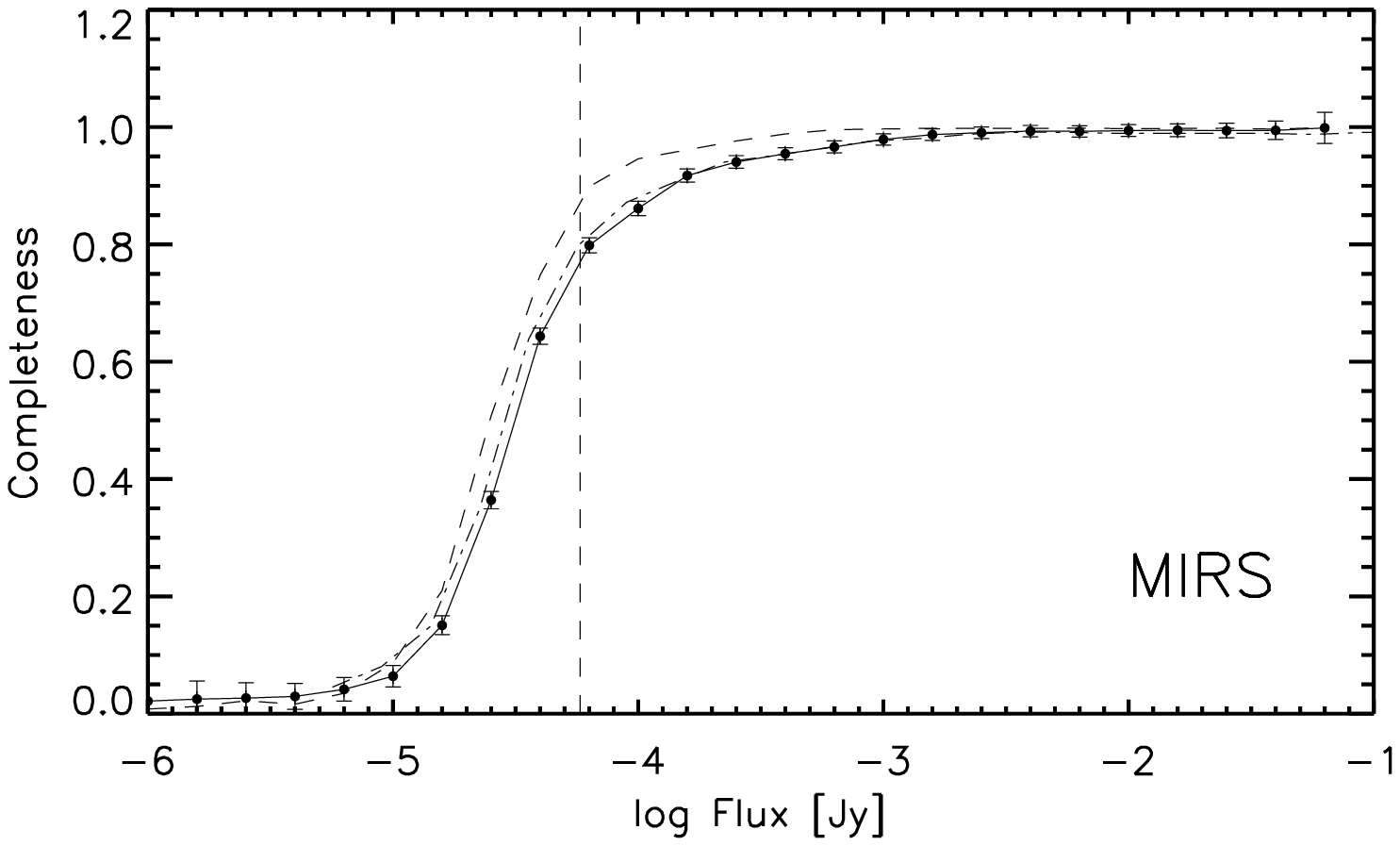}}
    \resizebox{8cm}{!}{\includegraphics{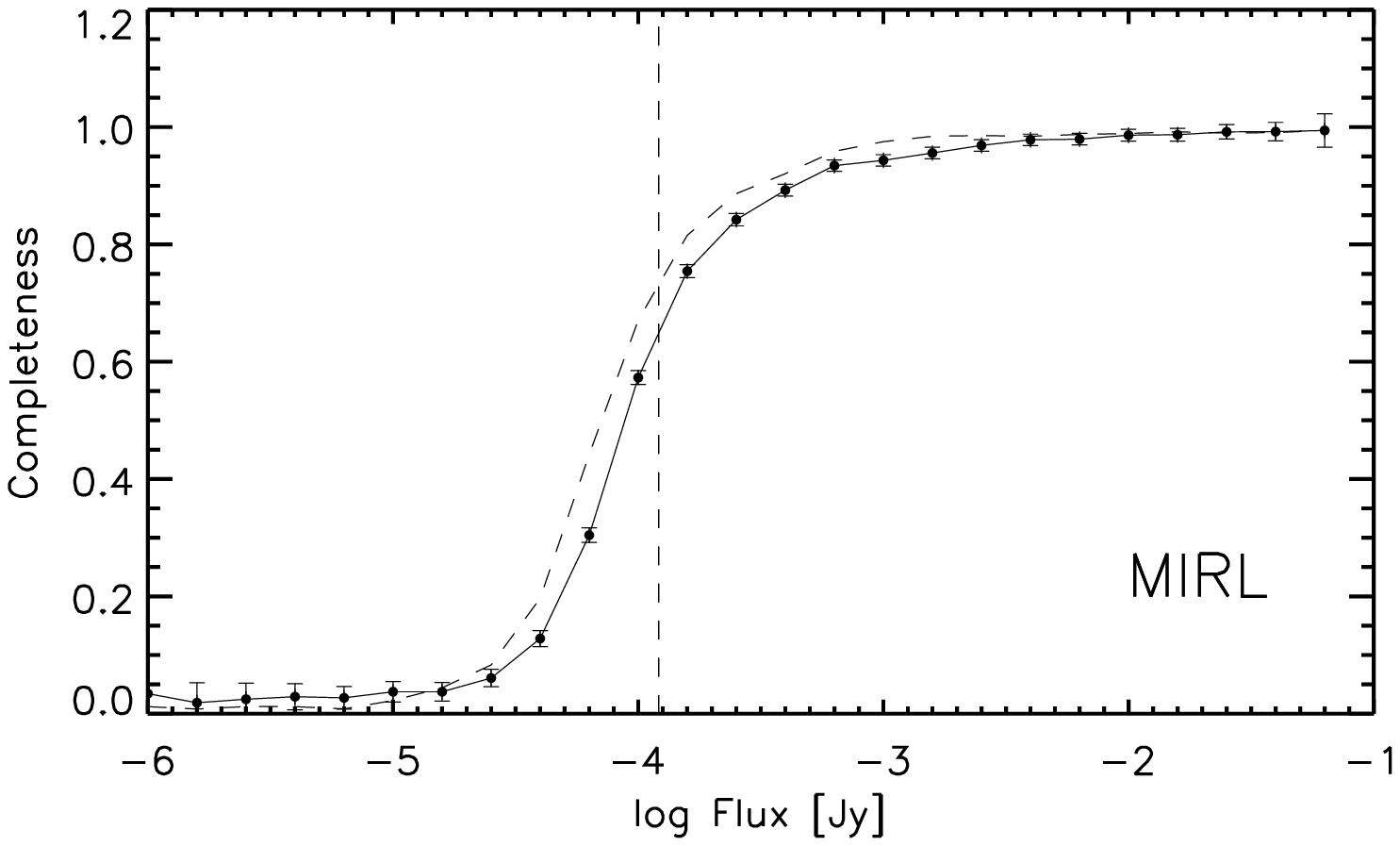}}
\end{center}
 \caption{Completeness simulation for the MIR-S and MIR-L channels. Solid and dashed lines represent the completeness calculated with the positive and negative images, respectively. For these cases, we assume that artificial sources have a flat spectrum in $f_\nu$.  
 The dot-dashed line in the upper panel indicates the completeness for red sources (see text in detail), where we adopted the average of 
 7, 9, and 11\,$\mu$m flux for red sources. The vertical dashed lines indicate 5\,$\sigma$ sensitivity in Wada et al. (2008) at S9W and L18W bands for MIR-S and MIR-L, respectively. 
 }
 \label{compl}
\end{figure}

We estimated the effect of source confusion in the completeness analysis as follows. We add artificial sources in the negative detection image, created by inverting the background subtracted detection image, and compare the results with the case for the normal positive image. In Figure \ref{compl}, we show the completeness thus obtained for both positive and negative images. For both MIR-S and MIR-L channels, we find that the completeness for the negative image is systematically higher than that for the positive image.
In the completeness analysis, we could wrongly identify real sources as artificial sources we generated. 
If such mis-identifications have significant impact on the completeness analysis, we would expect that the completeness found for the positive image with real sources would be higher than that for the negative image. Since this is not the case, the effect of mis-identification is likely to be negligible. The difference between the completeness for the positive image and for the negative image implies another important effect -- source blending. When we add artificial sources in the positive image, some of them would blend with real sources, which results in the positional shift of the extracted source and non-recovery. At the 5$\,\sigma$ sensitivity of IRC wide band images, i.e. S9W and L18W, the completeness for the negative image is $\sim$10\,\% higher than that for the positive image in both MIR-S and MIR-L channels.  This indicates that $\sim$10\,\% of mid-IR sources are missed, because of source blending. 

In order to evaluate the colour dependence of the completeness, we calculate the completeness of red sources as a function of the average flux of three MIR-S bands, i.e. $\langle f_\nu \rangle_\mathrm{MIR\-S} = 5.67\cdot f_{7\mu m}$ for the assumed colour, which is close to $f_{9\mu m}$. The resulting completeness is plotted in Figure \ref{compl}, and found to be close to the case for the flat spectrum. Thus, we conclude that the completeness has weak colour dependence, when the average fluxes of objects are considered. Since we consider very red sources in this simulation, it might be expected that the completeness of the longest wavebands, 11\,$\mu$m, using the multi-band detection image, could be worse than the case of using the $S11$ image as the detection image. The completeness of the single-band source extraction using the same image is reported by \cite{2008PASJ...60S.517W}. They obtained 80.7\,$\mu$Jy for the 50\,\% completeness at the $S11$ band. On the other hand, the 50\,\% completeness we obtained for red sources is 50.6\,$\mu$Jy at the same band, which is comparable to or even better than the value reported by \cite{2008PASJ...60S.517W}. Thus, on average, we see no sign of degraded completeness for red sources. 

We estimate the reliability of the catalogue as described below. Using the same method as for real sources, we made a source catalogue for the negative detection images for both MIR-S and MIR-L as described above. In band-merged catalogues, it is expected that sources detected in fewer bands would have lower reliability. In Table 3, we summarize the statistics of sources with the multiple band detection for 6 bands in MIR-S and MIR-L altogether. Although, from the negative image, we obtain some spurious sources with multiple-band detection, it turns out that almost all of them are rare cases produced by the effect of inverted real sources. Therefore, we do not take these spurious sources into account, and conclude that sources with multiple-band detection are highly reliable. In the negative images, there are 113 spurious sources with single band detection. These are all faint sources with the fluxes close to the detection limits. On the other hand, in the positive images, we find 2138 sources with single band detection. Given these numbers, we estimate that $\sim$5\,\% of sources with single-band detection are fake. This corresponds to 1.5\,\% of sources in the final catalogue.

\begin{table*}
\begin{center}
\caption{Summary of simulation for photometric errors and completeness}
\begin{tabular}{c|ccccccccc|cc}
\hline
\hline
Flux       &  \multicolumn{9}{c}{Photometric errors}   & \multicolumn{2}{c}{Completeness$^a$} \\
 (Jy)   &  N2  &   N3   &   N4 &   S7 &   S9W &   S11 &   L15 &   L18W &   L24 & MIR-S & MIR-L \\
\hline
3.0e-06  &   ...$^b$   &   ...   &  0.817  &   ...   &   ...   &   ...   &   ...   &   ...   &   ...   &  0.03  &  0.03  \\
6.0e-06  &   ...   &  0.678  &  0.395  &   ...   &   ...   &   ...   &   ...   &   ...   &   ...   &  0.04  &  0.03  \\
1.2e-05  &  0.351  &  0.326  &  0.284  &   ...   &   ...   &   ...   &   ...   &   ...   &   ...   &  0.10  &  0.04  \\
2.4e-05  &  0.230  &  0.152  &  0.139  &  0.971  &  1.056  &   ...   &   ...   &   ...   &   ...   &  0.34  &  0.06  \\
4.7e-05  &  0.151  &  0.085  &  0.088  &  0.600  &  0.652  &  0.834  &  1.674  &  1.694  &   ...   &  0.70  &  0.19  \\
9.4e-05  &  0.089  &  0.047  &  0.046  &  0.326  &  0.347  &  0.521  &  0.781  &  0.897  &   ...   &  0.85  &  0.54  \\
1.9e-04  &  0.055  &  0.031  &  0.024  &  0.172  &  0.183  &  0.264  &  0.427  &  0.456  &  1.009  &  0.93  &  0.79  \\
3.8e-04  &  0.027  &  0.015  &  0.011  &  0.087  &  0.094  &  0.143  &  0.244  &  0.261  &  0.504  &  0.95  &  0.89  \\
7.5e-04  &  0.015  &  0.008  &  0.006  &  0.044  &  0.047  &  0.070  &  0.121  &  0.137  &  0.255  &  0.97  &  0.94  \\
1.5e-03  &  0.007  &  0.004  &  0.003  &  0.022  &  0.024  &  0.034  &  0.066  &  0.069  &  0.134  &  0.99  &  0.95  \\
3.0e-03  &  0.004  &  0.002  &  0.002  &  0.011  &  0.012  &  0.017  &  0.035  &  0.036  &  0.068  &  0.99  &  0.97  \\
6.0e-03  &  0.002  &  0.001  &  0.001  &  0.006  &  0.007  &  0.009  &  0.018  &  0.017  &  0.035  &  0.99  &  0.98  \\
1.2e-02  &  0.001  &  0.001  &  0.000  &  0.003  &  0.003  &  0.004  &  0.010  &  0.010  &  0.019  &  0.99  &  0.99  \\
2.4e-02  &  0.001  &  0.000  &  0.000  &  0.002  &  0.002  &  0.002  &  0.005  &  0.005  &  0.010  &  0.99  &  0.99  \\
4.7e-02  &  0.000  &  0.000  &  0.000  &  0.001  &  0.001  &  0.001  &  0.002  &  0.002  &  0.005  &  1.00  &  0.99  \\
9.4e-02  &  0.000  &  0.000  &  0.000  &  0.001  &  0.000  &  0.001  &  0.001  &  0.002  &  0.003  &  1.00  &  1.00  \\
\hline 
\multicolumn{12}{l}{$^a$ Completeness in MIR-S and MIR-L detection images for the case of constant fluxes.} \\
\multicolumn{12}{l}{$^b$  Photometric errors are given if $|1-R_{ave}|< 0.2$, where $R_{ave}$ is the 
average of the input-to-output flux ratio}\\
\end{tabular}
\end{center}
\label{tb::error}
\end{table*}

\begin{table*}
\begin{center}
\caption{The number of sources with multiple MIR band detections}
\begin{tabular}{l|ccccccccc|cc}
\hline
\hline
\# of detected bands & 1 & 2 & 3 & 4 & 5 & 6 & Total \\
\hline
\# of sources & 2138 & 1736 & 1286 & 756 & 705 & 669 & 7284\\
Fraction & 0.29 & 0.24  & 0.17& 0.10& 0.097& 0.092 & 1\\
\hline 
\end{tabular}
\end{center}
\label{tb::multiband}
\end{table*}


\section{Catalogue}

\subsection{MIR-merged catalogue}
At this point, we had two catalogues, based on observations with the MIR-S and MIR-L channels, which we then merged together by eliminating duplicated entries. In order to choose the best photometric results from the duplicated entries, we checked the signal-to-noise ratio at each IRC band, and selected the source with the maximum number of $>3\,\sigma$ detections. 
Finally, we removed the sources whose fluxes are lower than $5\,\sigma$ in all MIR bands. We adopt the 5\,$\sigma$ sensitivity from Wada et al. (2008). We obtained 7284 mid-IR sources in the final catalogue.

The effects of source blending are not negligible in the process of catalogue generation, since we assume that sources with separations of $<3''$ are the same. There are multiple optical counterparts for 29\% (8\%) of MIR sources within $3''$ ($2''$). The ratio of source density indicates that only 3\% of optical sources in the Subaru/S-cam image are detectable in the MIR, and therefore the fraction of sources with serious blending in the IRC image should be much lower than the source fraction with multiple optical counterparts. 

For sources with moderate separation, inspection by eye can easily spot the problem of blending which affects both the source detection and photometry. In order to call attention to them, we set flags (group ID and the number of group member) to identify groups of MIR sources with separation of $<5''$. The sources with group ID should be treated with caution. Some of them have a common optical counterpart. 

A small portion of this band-merged catalogue is shown in Table 4, which is provided in full in the electric version of this paper.

\subsection{Optical identification}
\begin{figure}
\begin{center}
  \resizebox{8.5cm}{!}{\includegraphics{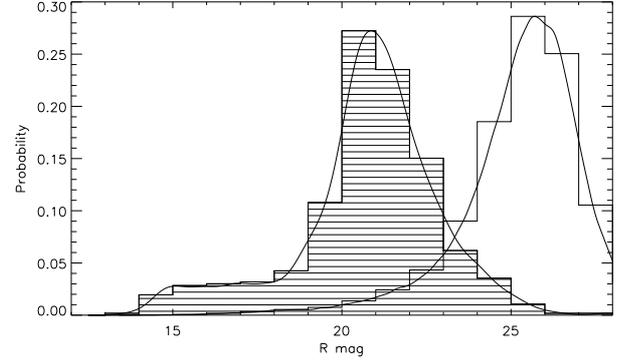}}
\end{center}
 \caption{Probability distribution function $q(m)$ (shaded histogram) and $n(m)$ (empty histogram). Probability densities estimated from these histograms are over-plotted with solid curves. 
}
 \label{LR}
\end{figure}

We made optical identifications within the Subaru/S-cam field. Although the S-cam field covers only 38\% of the NEP-Deep, it is deep enough to detect almost all of the {\it AKARI} MIR-detected sources. We found multiple optical counterparts for 29\,\% of MIR sources in our $R$-band S-cam image within 3$''$ radius. In order to find the best candidate, we adopt the maximum likelihood method (Sutherland \& Saunders 1992) for the optical identification. The likelihood ratio is defined as 
\begin{eqnarray}
L = \frac{q(m) f(x,y)}{n(m)}, 
\end{eqnarray}
where $q(m)$ is the infinitesimal probability that a MIR source has an optical magnitude of $m$, $f(x,y)$ is the probability distribution function of the positional error assumed to be a two-dimensional Gaussian, and $n(m)$ is the surface density of background objects with magnitude $m$. To derive the best estimate of $q(m)$, we defined a sub-sample of 1100 IRC all-band-detected sources, performed optical identification with a simple nearest neighbour method, and visually checked all of the optical identification. We derived $q(m)$ using this sub-sample with visually-confirmed optical identification. The resulting $q(m)$ is shown in Figure \ref{LR}, along with $n(m)$ from the Subaru $R$-band image. We assume that this distribution is not very different from that for general MIR sources detected in the NEP-Deep survey. For $f(x,y)$, we adopt the astrometric dispersions described above, i.e. 0.29$''$ for RA and 0.32$''$ for Dec. We calculated $L$ for objects within the search radius of 3$''$ and selected the object with the highest $L$ as the best candidate for the optical counterpart. A small portion of the resulting catalogue of optical counterparts is presented in Table 5.

As expected, the resulting optical ID with likelihood ratio is sometimes different from the nearest neighbour. For 915 MIR sources with multiple optical counterparts 
in the S-cam field, we found that 234 sources have an optical ID which is not the nearest neighbour. For these sources, we visually inspected the results, but it was not useful to identify correct counterparts. Thus, we have serious problems of optical ID for $\sim 10$\% of MIR sources. Also, we spotted several pairs of MIR sources which happen to have a common optical counterpart. Therefore, optical identification of sources with close neighbours should be treated with caution. In the catalogue, these sources are identified with the group ID flag. 

\begin{figure}
\begin{center}
  \resizebox{8.5cm}{!}{\includegraphics{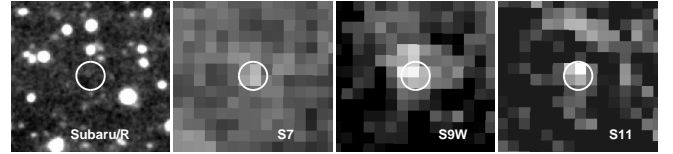}}
\end{center}
 \caption{Postage stamp of a mid-IR source (MIRS4404) with no obvious optical counterparts in the Subaru/S-cam image. Circles indicate the IRC/MIR-S position with 3$''$ radius. 
}
 \label{noID}
\end{figure}

Out of 7284 mid-IR sources, we found 3162 sources in the S-cam field and optical counterparts for all but 79 sources. Most of mid-IR sources with no optical ID were detected with only a single mid-IR band, which means they are likely to be unreliable sources. However, we find that 9 sources out of 79 were detected in more than 3 mid-IR bands. As a result of this visual inspection, we found that their optical counterparts are pairs of bright sources or lie close to optically bright sources affected by spikes, which hamper the correct identification. Interestingly, we find a genuine optically-blank mid-IR source even with the deep Subaru/S-cam image, which is shown in Figure \ref{noID}. The $R-L15$ colour of this source is 9.5 mag (AB), which is about 3\,mag redder than the criterion to choose the reddest infrared sources, such as faint dust-obscured galaxies \citep[DOGs --][]{2008ApJ...677..943D}.

\subsection{Star-galaxy separation}

\begin{figure}
\begin{center}
  \resizebox{8.5cm}{!}{\includegraphics{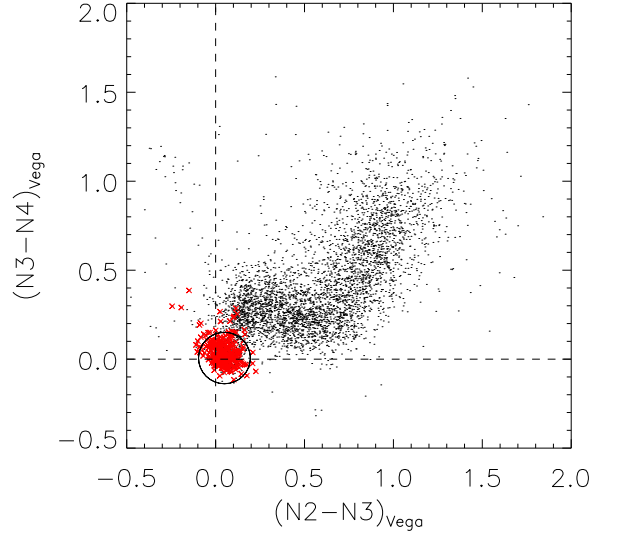}}
  \end{center}
 \caption{Near-IR colour-colour plot in Vega magnitudes for star-galaxy separation. Objects with large stellarity ($>0.9998$) are indicated with crosses. A large circle indicates the adopted boundary for star-galaxy sepation (see text in detail). Zero-colours are marked with dashed lines. 
}
 \label{sg}
\end{figure}

We mainly use $N2-N3$ and $N3-N4$ colours for identification of stars, since these near-IR bands are the most sensitive IRC bands. In the Vega-based magnitude system, the near-IR colours of normal stars are close to zero. In this subsection, all magnitudes are given in the Vega-based magnitudes. We firstly made a tentative list of stars, based on a stellarity 
measured in the CFHT $r'$-band image \citep{2007ApJS..172..583H}. In Figure \ref{sg}, we show the $N3-N4$ versus $N2-N3$ colour-colour plot for star-galaxy separation. In this plot, objects with a large stellarity, i.e. stellar objects, make a clump around zero-colors, as expected. The average colours of these stellar objects are $\langle N2-N3 \rangle = 0.049\pm 0.057$ and $\langle N3-N4 \rangle = 0.006\pm 0.056$\footnote{With no colour corrections}. 
We draw a circular boundary in this colour-colour plot to separate stars and galaxies defined with the radius $\Delta C (\equiv \sqrt{\Delta(N2-N3)^2 + \Delta(N3-N4)^2} )$ of 0.15, where we adopt the mean colours of stellar objects as a centre. This classification results in 673 stars, out of 7284 sources. 

A large $\Delta C$ for the colour boundary would increase the completeness of stars, but could cause mis-classifications of nearby galaxies as stars. Therefore, we additionally use following criteria;  $-1<N2-S11<1$ and $N2<13$~mag for stars. This colour is useful to separate stars without mid-IR excess from nearby galaxies. 

Furthermore, some bright stars were saturated at the central pixels, specifically in the near-IR bands. This causes yet another mis-classification if we use near-IR bands for star-galaxy separation. To remedy this effect, we adopt additional criteria for stars, i.e. $S7 < 10$~mag and $-1<S7-S11<1$. With these criteria, we finally obtained 720 stars in total.  

There is a prominent galactic source in the NEP-Deep region -- a planetary nebula NGC6543. The IRC detected filaments of this planetary nebula, which should be flagged. Inside this PN area, photometry of distant sources is severely affected by these filaments. Therefore, we simply flagged out all of sources within the PN area. The central position of a circular mask is adjusted to encompass the PN area with the radius of 194$''$.

\section{Discussion}

\begin{figure}
\begin{center}
  \resizebox{8cm}{!}{\includegraphics{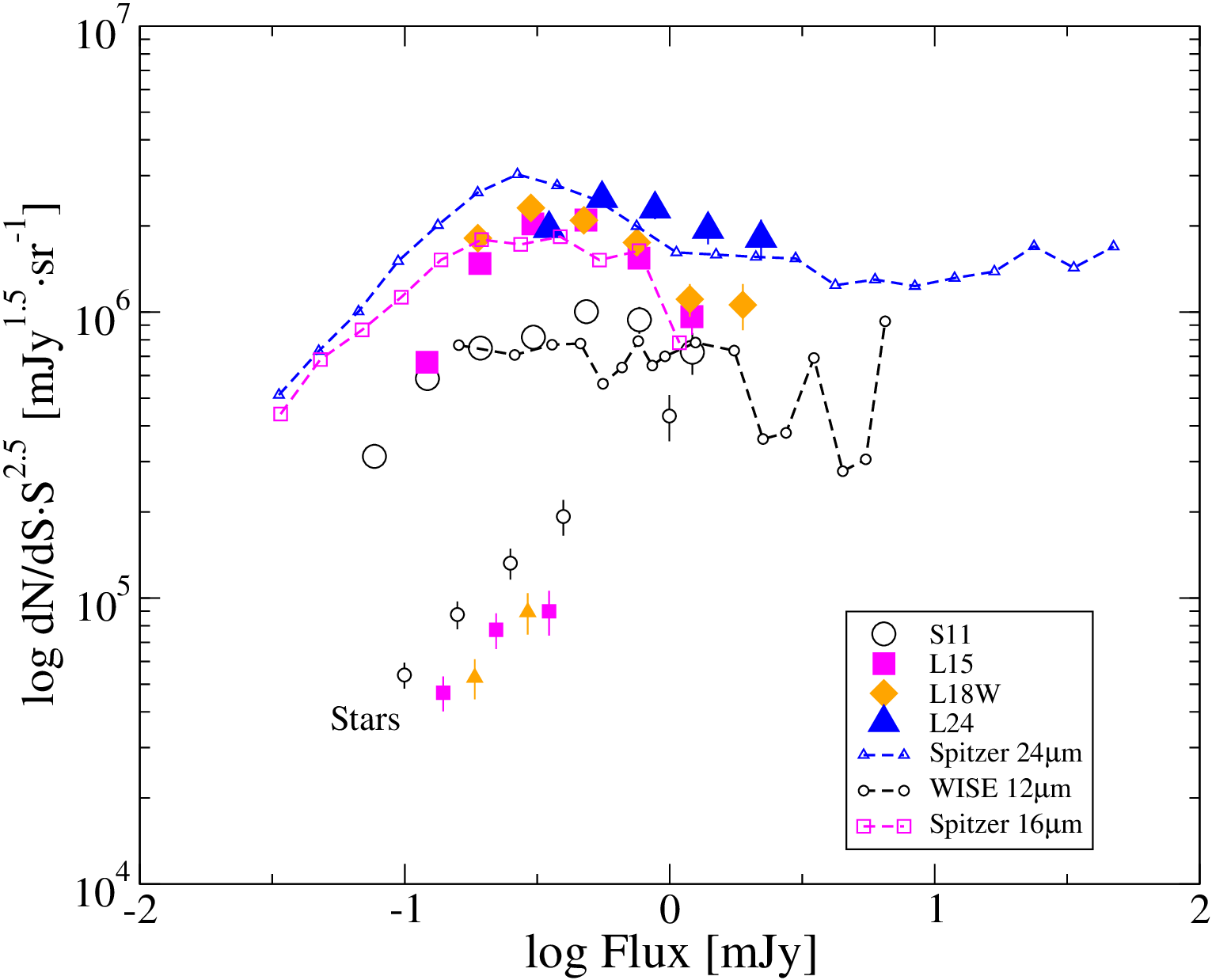}}
  \end{center}
 \caption{Number counts of galaxies and stars at 11, 15, 18, and 24\,$\mu$m. Large and small symbols with error bars indicate galaxy and star counts, respectively. See legend for individual bands. For the L24 band, the number of stars is not enough to derive statistical counts. Dashed lines with symbols indicate Spitzer 24\,$\mu$m counts \citep{2004ApJS..154...70P}, Spitzer 16\,$\mu$m counts \citep{2011AJ....141....1T}, and WISE 12\,$\mu$m counts \citep{2011ApJ...735..112J} from top to bottom. We show the number counts for flux bins containing at least 30 objects and with flux greater than the 5\,$\sigma$ detection limit. All counts are corrected for completeness. 
}
 \label{count}
\end{figure}

We calculated the number counts of mid-IR sources at 11, 15, 18 and 24\,$\mu$m shown in Figure \ref{count}. In deriving the counts, we first defined the area to be used for each image based on the weight map. We found most low-weight areas in the north-east part of the field, in which {\it AKARI} observations suffered from the Earth shine. To reduce this effect, we excluded the field with the weight of the lowest 20\,\%. As a result of this cut, we ended up with areas of 0.46, 0.53, 0.51, and 0.46\,deg$^2$ for 11, 15, 18, and 24\,$\mu$m, respectively. The excluded area is 0.167\,deg$^2$ at most. The counts are corrected for the completeness. We compare our $L24$ counts with {\it Spitzer} 24\,$\mu$m counts \citep{2004ApJS..154...70P}, and find a reasonable agreement. However, even with the completeness correction, the counts below the 5\,$\sigma$ detection limit exhibit a significant under-estimation, compared to the {\it Spitzer}'s counts. Therefore, in Figure \ref{count}, we only show the counts above the $5\,\sigma$ detection limits. For other mid-IR number counts in {\it AKARI} bands, see \cite{2007PASJ...59S.515W, 2008PASJ...60S.517W, 2009PASJ...61..375L,2010A&A...514A...8P} and Pearson et al. (2011, in preparation). 

The wavelength-dependence of number counts from 11-to-24\,$\mu$m could be explained by the SEDs of galaxies, where PAHs and hot dust emission dominate. We find the least number of sources at 11\,$\mu$m, where the prominent PAH emission around 8\,$\mu$m is redshifted out of the bandpass for $z\ga 0.4$. 

\begin{figure}
\begin{center}
  \resizebox{8cm}{!}{\includegraphics{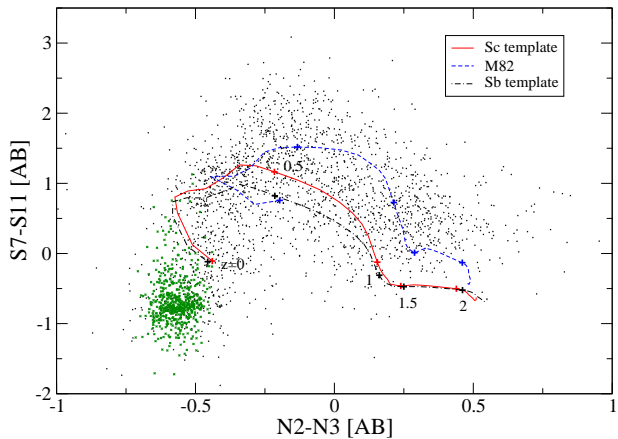}}
  \end{center}
 \caption{Colour-colour plot with {\it AKARI} bands, using $N2, N3, S7$ and $S11$. Dots (black) and small crosses (green) indicate 
 galaxies and stars, respectively, detected at more than 3\,$\sigma$ in all of 4 bands. Solid line (red) indicates 
 the colour of Sc galaxy template as a function of redshift. Dot-dashed (black) and dashed (blue) lines are for Sb 
 galaxy and M82 template, respectively. The SED templates are taken from \cite{2007ApJ...663...81P}. 
 }
 \label{colcol}
\end{figure}

A typical redshift of catalogued galaxies can be estimated from {\it AKARI} colours. Figure \ref{colcol} shows the colour-colour plot using {\it AKARI} bands, $S7-S11$ versus $N2-N3$. Both colours have a good dynamic range specifically at $z<1$, owing to the 1.6\,$\mu$m bump of the stellar emission and the PAH emission at $8\,\mu$m. At $z<1$, $N2-N3$ almost continuously increases with increasing redshift, because of the 1.6\,$\mu$m bump. On the other hand, $S7-S11$ has a maximum at $z\sim 0.5$, since the redshifted PAH 8\,$\mu$m feature is captured by $S11$ bands at that redshift. These trends explain the arch-shaped distribution of galaxies in this colour-colour plot. From this colour-colour plot alone, it is safe to conclude that most of {\it AKARI} mid-IR sources in the NEP-Deep field lie at $z<1$. The reddest $S7-S11$ galaxies may have the strongest PAH emission features at $z\sim 0.5$, and are studied in detail by \cite{2010A&A...514A...5T} as `PAH-selected' galaxies.

\section{Summary} 
We have generated a mid-IR source catalogue based on images from the {\it AKARI} NEP-Deep survey presented by \cite{2008PASJ...60S.517W}. This catalogue was designed to include most of the sources detected at 7, 9, 11, 15, or 18\,$\mu$m. In the catalogue we report 7284 sources in the nearly circular area of the NEP-Deep, covering 0.67\,deg$^2$. From the simulation of artificial sources, we have estimated the photometric errors in each band and the  completeness in both MIR-S and MIR-L channels. The star-galaxy separation is based solely on {\it AKARI} photometry. For sources in the Subaru/S-cam field, where the optical photometry is deep enough to detect most of {\it AKARI} mid-IR sources, we performed optical identification with the likelihood ratio method. The mid-IR number counts we derived show a drastic increase of sources between 11\,$\mu$m and 15\,$\mu$m, owing to the effect of redshifted PAH emission. At the flux level of 1\,mJy, we found the highest source density per flux bin in the 24\,$\mu$m band. Based on redshift-sensitive colours, $N2-N3$ and $S7-S11$, most of {\it AKARI} mid-IR sources are found to lie at $z\la 1$. 


\section*{Acknowledgements} 
We would like to thank all the AKARI team members for their extensive 
efforts. 
This work is supported by the Japan Society for the Promotion 
of Science (JSPS; grant number 18$\cdot$7747 and 21340042). 
H.M. Lee was supported by NRF grant No. 2006-341-C00018.
MI was suppored by the grant No. 2010-0000712 of the NRFK/MEST.
This research is based on observations with {\it AKARI}, a JAXA project with the participation of ESA, 
data collected at Subaru Telescope, which is operated by the National Astronomical Observatory of Japan, 
observations at Kitt Peak National Observatory, National Optical Astronomy Observatory, and also 
observations obtained with MegaPrime/MegaCam, a joint project of CFHT and CEA/DAPNIA. 
The authors wish to thank the referee, whose comments are helpful to improve the contents of this work. 
\bibliography{reference}

\begin{landscape}

\begin{table}
\begin{minipage}{230mm}
\caption{Mid-IR source catalogue of the {\it AKARI} NEP-Deep field }
\tiny
\tiny
\begin{tabular}{lccccccccccccccccccccc}
\hline
ID & Name & RA & Dec & 
$N2$ &
$N3$ &
$N4$ &
$S7$ &
$S9W$ &
$S11$ &
$L15$ &
$L18W$ &
$L24$ &
Star & Group ID \\
&  &  \multicolumn{2}{c}{[J2000.0]}  &
$f_\nu$ & 
$f_\nu$ & 
$f_\nu$ & 
$f_\nu$ & 
$f_\nu$ & 
$f_\nu$ & 
$f_\nu$ & 
$f_\nu$ & 
$f_\nu$ &   &  \\ 
& AKARI-NEPD& & &
Jy & 
Jy & 
Jy & 
Jy & 
Jy & 
Jy & 
Jy & 
Jy & 
Jy &  \\ 
(1) & (2) & (3) & (4)  & (5) & (6) & (7) & (8) &
(9) & (10) & (11) & (12) & (13)  & (14) & (15)  \\
\hline
... & & & & & & & & & & & & & & \\
MIRS464  &  J175651.47+661445.1 & 269.21447 &  66.24587 &  1.42e-05 &  1.97e-05 &  2.16e-05 &  4.83e-06 &  3.65e-05 &  7.28e-05 &  0.00e+00 &  9.27e-05 &  1.62e-04 &  0 &   0  \\
MIRS465  &  J175538.70+661455.6 & 268.91126 &  66.24878 &  2.95e-05 &  2.62e-05 &  2.06e-05 &  7.06e-05 &  7.86e-05 &  1.48e-04 &  0.00e+00 &  0.00e+00 &  0.00e+00 &  0 &   0  \\
MIRS466  &  J175724.90+661425.9 & 269.35379 &  66.24055 &  1.01e-03 &  5.22e-04 &  3.35e-04 &  1.89e-04 &  1.46e-04 &  1.87e-04 &  3.28e-04 &  2.96e-04 &  4.04e-04 &  1 &   0  \\
MIRS467  &  J175712.36+661442.4 & 269.30152 &  66.24512 &  7.83e-05 &  6.28e-05 &  4.29e-05 &  4.51e-05 &  1.00e-04 &  1.51e-04 &  2.74e-04 &  3.24e-04 &  3.10e-04 &  0 &   0  \\
MIRS468  &  J175457.95+661440.2 & 268.74146 &  66.24452 &  7.77e-04 &  4.55e-04 &  2.55e-04 &  1.55e-04 &  1.04e-04 &  7.49e-05 &  0.00e+00 &  6.96e-05 &  1.84e-04 &  2 &   0  \\
MIRS469  &  J175524.63+661452.7 & 268.85266 &  66.24798 &  0.00e+00 &  0.00e+00 &  0.00e+00 &  0.00e+00 &  5.16e-05 &  7.28e-05 &  7.34e-05 &  0.00e+00 &  0.00e+00 &  0 &   0  \\
MIRS471  &  J175629.97+661436.9 & 269.12488 &  66.24360 &  6.27e-05 &  3.95e-05 &  2.93e-05 &  1.78e-05 &  6.86e-05 &  1.52e-05 &  3.13e-05 &  6.52e-05 & -1.00e+00 &  0 &   0  \\
MIRS474  &  J175801.17+661506.9 & 269.50489 &  66.25192 &  0.00e+00 &  0.00e+00 &  0.00e+00 &  0.00e+00 &  0.00e+00 &  0.00e+00 &  1.24e-04 &  9.64e-05 &  4.29e-05 &  0 &   0  \\
MIRS475  &  J175513.85+661417.6 & 268.80771 &  66.23823 &  1.69e-02 &  9.98e-03 &  5.26e-03 &  2.30e-03 &  1.74e-03 &  1.11e-03 &  4.97e-04 &  3.28e-04 &  8.84e-05 &  2 &   0  \\
MIRS477  &  J175718.83+661427.3 & 269.32849 &  66.24094 &  0.00e+00 &  0.00e+00 &  0.00e+00 &  0.00e+00 &  4.65e-05 &  0.00e+00 &  1.24e-05 &  1.47e-04 &  2.63e-04 &  0 &   7  \\
MIRS478  &  J175721.29+661452.5 & 269.33871 &  66.24794 &  1.07e-05 &  1.62e-05 &  1.25e-05 &  1.32e-05 &  0.00e+00 &  0.00e+00 &  2.08e-04 &  2.36e-04 &  3.59e-04 &  0 &   8  \\
... & & & & & & & & & & & & & & \\
\hline 
\\
\end{tabular}
Notes.--- See the electronic version of this paper for the complete catalogue. {\it Flux errors are tabulated only in the electronic version}. Column~(1): AKARI mid-IR source ID. Column~(2): Source Name. Columns~(3) and (4): AKARI/IRC J2000.0 RA and Dec. Columns~(5) through (13): IRC flux densities. Undetected sources have a null value, while -1 means the out-of-field. Column~(14): Star/galaxy flag [0: galaxies, 1: stars selected with near-IR colours, 2: bright stars identified with mid-IR photometry, 3: sources near the planetary nebula NGC6543]. Column~(15): Group ID for close neighbours within 5$''$ radius. 
\end{minipage}
\end{table}

\begin{table}
\begin{minipage}{230mm}
\caption{Photometry from UV to NIR with CFHT/Megacam, Subaru/S-cam and KPNO2.1m/FLAMINGOS}
\tiny
\tiny
\begin{tabular}{lccccccccccccccccccccc}
\hline
ID & RA & Dec & 
$\Delta \theta$ & 
$u^*$ & 
$\Delta u^*$& 
$B$ & 
$\Delta B$& 
$V$ & 
$\Delta V$& 
$R$ & 
$\Delta R$& 
$i'$ & 
$\Delta i'$& 
$z'$ & 
$\Delta z'$& 
$J$ & 
$\Delta J$& 
$K_s$ & 
$\Delta K_s$& \# of Neighb.
\\
&    \multicolumn{2}{c}{Optical / [J2000.0]}  & arcsec &
mag & mag &
mag & mag &
mag & mag &
mag & mag &
mag & mag &
mag & mag &
mag & mag &
mag & mag & 
\\ 
(1) & (2) & (3) & (4)  & (5) & (6) & (7) & (8) &
(9) & (10) & (11) & (12) & (13)  & (14) & (15)  & (16) & (17) & (18) & (19) & (20) & (21) \\
\hline
... & & & & & & & & & & & & & & \\
MIRS1596 & 268.44467 &  66.39353 &  0.89 &   24.33 &    0.07 &   22.80 &    0.01 &   21.48 &    0.01 &   20.62 &    0.01 &   19.79 &    0.01 &   19.25 &    0.01 &   99.00 &   99.00 &   99.00 &   99.00 &  2 \\
MIRS1597 & 268.35163 &  66.39347 &  0.27 &   23.74 &    0.05 &   22.84 &    0.01 &   21.65 &    0.01 &   21.04 &    0.01 &   20.60 &    0.01 &   20.17 &    0.01 &   99.00 &   99.00 &   99.00 &   99.00 &  1 \\
MIRS1598 & 269.13741 &  66.39712 &  0.10 &   22.78 &    0.01 &   21.97 &    0.01 &   20.84 &    0.01 &   20.23 &    0.01 &   19.87 &    0.01 &   19.54 &    0.01 &   18.94 &    0.03 &   18.46 &    0.03 &  1 \\
MIRS1599 & 268.94945 &  66.39747 &  0.06 &   99.00 &   99.00 &   24.23 &    0.01 &   23.13 &    0.01 &   22.05 &    0.01 &   21.39 &    0.01 &   20.97 &    0.01 &   20.12 &    0.07 &   19.27 &    0.06 &  1 \\
MIRS1600 & 268.73970 &  66.39690 &  0.26 &   23.25 &    0.02 &   22.75 &    0.01 &   22.24 &    0.01 &   21.49 &    0.01 &   21.23 &    0.01 &   20.99 &    0.01 &   20.72 &    0.10 &   19.94 &    0.09 &  1 \\
MIRS1607 & 268.80148 &  66.39745 &  0.18 &   23.74 &    0.03 &   22.87 &    0.01 &   21.81 &    0.01 &   21.17 &    0.01 &   20.86 &    0.01 &   20.44 &    0.01 &   20.10 &    0.07 &   19.22 &    0.07 &  1 \\
MIRS1608 & 268.99580 &  66.39949 &  0.12 &   22.63 &    0.02 &   21.62 &    0.01 &   20.31 &    0.01 &   19.65 &    0.01 &   19.31 &    0.01 &   18.99 &    0.01 &   18.41 &    0.02 &   17.99 &    0.02 &  1 \\
MIRS1613 & 268.68909 &  66.40347 &  1.95 &   22.87 &    0.02 &   19.07 &    0.01 &   18.65 &    0.01 &   18.41 &    0.01 &   18.83 &    0.01 &   18.12 &    0.01 &   18.08 &    0.04 &   18.56 &    0.03 &  2 \\
MIRS1614 & 268.97998 &  66.39938 &  0.13 &   22.17 &    0.01 &   21.70 &    0.01 &   20.91 &    0.01 &   20.64 &    0.01 &   20.52 &    0.01 &   20.36 &    0.01 &   20.25 &    0.08 &   20.18 &    0.14 &  1 \\
MIRS1615 & 268.66240 &  66.38812 &  0.44 &   19.23 &    0.01 &   18.52 &    0.01 &   17.82 &    0.01 &   17.36 &    0.01 &   17.17 &    0.01 &   16.80 &    0.01 &   16.20 &    0.01 &   15.90 &    0.01 &  1 \\
... & & & & & & & & & & & & & & \\
\hline 
\\
\end{tabular}
Notes.--- This table includes a subset of MIR sources which are detected with Subaru/S-cam, covering part of the NEP-Deep field. See the electronic version of this paper for the complete catalogue. Column~(1): AKARI mid-IR source ID. Columns~(2) and (3): Subaru/S-cam J2000.0 RA and Dec. Column~(4): Angular separation between mid-IR and optical position.  Columns~(5) through (20): UV--Optical--NIR magnitudes and errors in AB. Undetected sources have entries ``99". Minimum errors are assumed to be 0.01. Not corrected for the galactic extinction. Column~(19): Number of optical neighbours within 3$''$ radius in the S-cam catalogue. 
\end{minipage}
\end{table}

\end{landscape}

\end{document}